\documentclass[preprints,letter,accept,oneauthor,pdftex]{Definitions/mdpi} 

\firstpage{1}
\usepackage{graphicx}
\usepackage{bm,amsfonts,amsthm,placeins}

\pubvolume{00}
\issuenum{0}
\articlenumber{5}
\pubyear{2019}
\copyrightyear{2019}
\history{Received: date; Accepted: date; Published: date}

\Title{Wannier states of fcc symmetry qualifying paramagnetic NiO to
  be a Mott insulator}

\Author{Ekkehard Kr\"uger \orcidA{}}

\AuthorNames{Firstname Lastname, Firstname Lastname and Firstname Lastname}

\address[1]{%
Institut f{\"u}r Materialwissenschaft, Materialphysik,
 Universit{\"a}t Stuttgart, D-70569 Stuttgart, Germany; ekkehard.krueger@imw.uni-stuttgart.de}

\corres{Correspondence: ekkehard.krueger@imw.uni-stuttgart.de}

\abstract{This letter extends my recent paper on antiferromagnetic NiO
  [Structural Distortion Stabilizing the Antiferromagnetic and
  Insulating Ground State of NiO, Symmetry 2020, 12(1), 56] by
  including also the paramagnetic phase of this compound. I report
  evidence that paramagnetic NiO possesses a narrow, roughly
  half-filled energy band that produces a nonadiabatic atomic-like
  motion providing the basis for a Mott insulator in the paramagnetic
  phase. While the atomic-like motion operating in the
  antiferromagnetic phase is adapted to the symmetry of the
  antiferromagnetic state, in the paramagnetic phase the related
  localized states are represented by optimally localized Wannier
  functions possessing the full fcc symmetry of paramagnetic NiO. The
  nonadiabatic Wannier states are twofold degenerate, have $d$-like
  symmetry and are situated at the Ni atoms.}

\begin{document}


\section{Introduction}
In my previous paper~\cite{enio} on NiO, I reported evidence that the
structural distortion of antiferromagnetic NiO stabilizes the
antiferromagnetic and insulating ground state of this compound. This
state is enabled by the nonadiabatic atomic-like motion of the
electrons in the ``magnetic super band'' defined within the
nonadiabatic Heisenberg model (NHM)~\cite{enio}. Because this
atomic-like motion only exists in the space group of the
antiferromagnetic phase, paramagnetic NiO should be metallic.

However, also paramagnetic NiO is found to be an
insulator~\cite{brandow}. As is generally accepted~\cite{trimarchi},
Mott insulation is in any case a manifestation of atomic-like
electrons~\cite{hubbard} occupying localized orbitals as long as
possible and performing their band motion by hopping from one atom to
another. In view of my observation, that the nonadiabatic atomic-like
motion defined within the NHM is evidently responsible for the
insulating ground states of antiferromagnetic BaMn$_2$As$_2$ and
antiferromagnetic NiO~\cite{enio}, I ask in this letter whether also
in the conventional band structure of the fcc paramagnetic phase of
NiO there exists a narrow half-filled energy band producing a
nonadiabatic atomic-like motion at the Fermi level. In the following
Section~\ref{sec:wf} I show that this is actually the case.

\section{Nonadiabatic Atomic-like motion in Paramagnetic NiO}
\label{sec:wf}
Any atomic-like motion as defined within the NHM is represented by
optimally localized symmetry-adapted Wannier function being an exact
unitary transformation of the Bloch functions of a narrow, roughly
half-filled energy band~\cite{enio}. For the construction of such
Wannier functions we need a closed energy band in the conventional
band structure~\cite{enio} not connected by symmetry with other
bands. Just as in my former papers, a closed complex of $\mu$
individual energy bands is referred to a closed (single) band
consisting of $\mu$ branches~\cite{theoriewf}. However, such narrow
closed bands generally cannot be found in the band structures of the
metals unless we consider superconducting or magnetic
bands~\cite{theoriewf} possessing a reduced symmetry.


 \begin{figure}[H]
 \centering
 \includegraphics[width=.7\textwidth,angle=0]{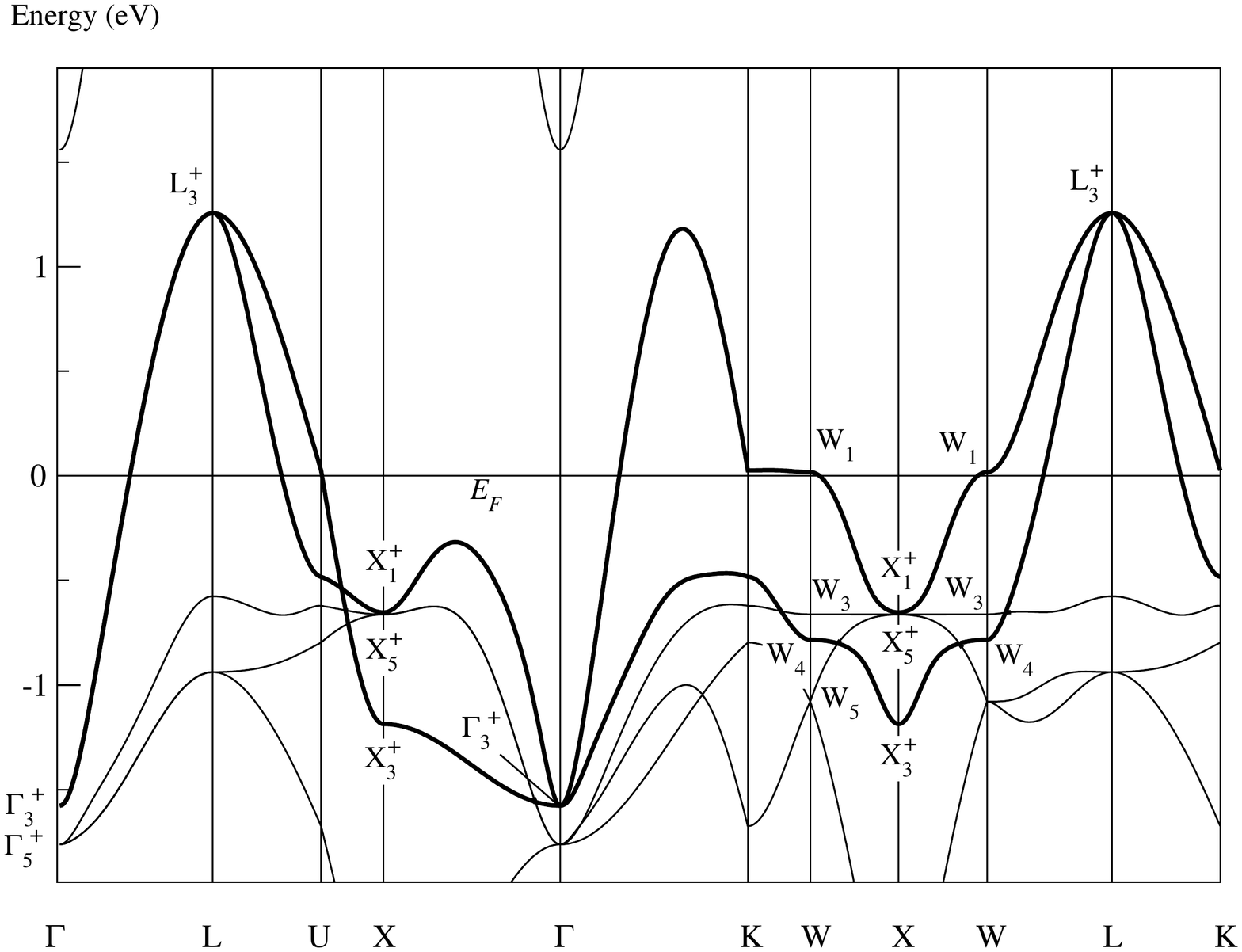}%
 \caption{Conventional band structure of paramagnetic fcc
   NiO as given in Figure 1 of~\cite{enio}. The notations of the
   points of symmetry in the Brillouin zone for $\Gamma^f_c$ follow
   Figure 3.14 of \cite{bc}, and the symmetry labels are defined in
   Table A1 of~\cite{enio}. The ``insulating band'' is highlighted by
   the bold line.  }
 \label{fig:bs}
 \end{figure}


 An exception (but presumably not the only exception) is paramagnetic
 NiO, as demonstrated by Figure~\ref{fig:bs} and
 Table~\ref{tab:wf_225}. Figure~\ref{fig:bs} shows the conventional
 band structure of paramagnetic NiO as determined in~\cite{enio}. Each
 row in Table~\ref{tab:wf_225} defines a band with Bloch functions
 that can be unitarily transformed into Wannier functions
 being 
\begin{itemize}
\item as well localized as possible (according to Definition 5 of~\cite{theoriewf}); 
\item centered at the Ni (Table a) or O (Table b) atoms; and
\item symmetry-adapted to $Fm3m$ (according to Equation (10) of~\cite{theoriewf}).
\end{itemize}
Thus, the Wannier functions of each listed band may define a
nonadiabatic atomic-like motion in paramagnetic NiO when the band is
roughly half-filled~\cite{enhm}.

The bands in Table~\ref{tab:wf_225} are determined following Theorem 5
of Ref.\ \cite{theoriewf}. However, the group theory of Wannier
functions presented in~\cite{theoriewf} is consistently restricted to
Wannier functions belonging to a one-dimensional representation of the
point group of their position $G_{0p}$, as it is suggested by
superconducting and magnetic bands (Section 2.2
of~\cite{theoriewf}). Nevertheless, the formalism can be extended to
degenerate Wannier functions. This shows that Theorem 5 is applicable
to degenerate Wannier functions when the complex numbers
$d_i(\alpha )$ in Equation (39) of~\cite{theoriewf} are interpreted as
the characters $\chi_i(\alpha )$ of the $i$th representation of
$G_{0p}$. Clearly, the $d_i(\alpha )$ and $\chi_i(\alpha )$ are
identical for one-dimensional representations.

 By inspection of Figure~\ref{fig:bs} and Table~\ref{tab:wf_225} we
 recognize that, first, the ``active band'' highlighted by the bold
 lines in the band structure of paramagnetic NiO has the symmetry of
 band 5 in Table~\ref{tab:wf_225} a and that, secondly, no partly
 filled band in the band structure of paramagnetic NiO possesses the
 symmetry of any band in Table~\ref{tab:wf_225} b. Thus, the active
 band provides localized states in the paramagnetic phase of NiO
 allowing the electrons to perform a strongly correlated nonadiabatic
 atomic-like motion stabilized by the nonadiabatic condensation energy
 $\Delta E$~\cite{enhm}. The nonadiabatic localized states are
 situated at the Ni atoms and possess $\Gamma^+_3 (E_g)$ symmetry
 ensuring that the nonadiabatic Hamiltonian of the atomic-like
 electrons commutes with the symmetry operators of the cubic space
 group $Fm3m$~\cite{enhm}.


\begin{table}[H]
\caption{
Symmetry labels of the Bloch functions at the points of symmetry in
the Brillouin zone for $Fm3m$ (225) of all
the energy bands with symmetry-adapted and optimally  
localized Wannier functions centered at the Ni (Table a) and O
(Table b) atoms, respectively. 
\label{tab:wf_225}}
\begin{center}
\begin{tabular}[t]{cccccc}
{\bf (a)} & Ni($000$) & $\Gamma$ & $X$ & $L$ & $W$\\
\hline
Band 1 & $\Gamma^+_1$ & $\Gamma^+_1$ & $X^+_1$ & $L^+_1$ & $W_1$\\
Band 2 & $\Gamma^+_2$ & $\Gamma^+_2$ & $X^+_3$ & $L^+_2$ & $W_4$\\
Band 3 & $\Gamma^-_2$ & $\Gamma^-_2$ & $X^-_3$ & $L^-_2$ & $W_2$\\
Band 4 & $\Gamma^-_1$ & $\Gamma^-_1$ & $X^-_1$ & $L^-_1$ & $W_3$\\
Band 5 & $\Gamma^+_3$ & $\Gamma^+_3$ & $X^+_1$ + $X^+_3$ & $L^+_3$ & $W_1$ + $W_4$\\
Band 6 & $\Gamma^-_3$ & $\Gamma^-_3$ & $X^-_1$ + $X^-_3$ & $L^-_3$ & $W_2$ + $W_3$\\
Band 7 & $\Gamma^+_4$ & $\Gamma^+_4$ & $X^+_2$ + $X^+_5$ & $L^+_2$ + $L^+_3$ & $W_2$ + $W_5$\\
Band 8 & $\Gamma^+_5$ & $\Gamma^+_5$ & $X^+_4$ + $X^+_5$ & $L^+_1$ + $L^+_3$ & $W_3$ + $W_5$\\
Band 9 & $\Gamma^-_4$ & $\Gamma^-_4$ & $X^-_2$ + $X^-_5$ & $L^-_2$ + $L^-_3$ & $W_4$ + $W_5$\\
Band 10 & $\Gamma^-_5$ & $\Gamma^-_5$ & $X^-_4$ + $X^-_5$ & $L^-_1$ + $L^-_3$ & $W_1$ + $W_5$\\
\hline\\
\end{tabular}


\begin{tabular}[t]{cccccc}
\\
{\bf (b)} & O($\overline{\frac{1}{2}}\frac{1}{2}\frac{1}{2}$) & $\Gamma$ & $X$ & $L$ & $W$\\
\hline
Band 1 & $\Gamma^+_1$ & $\Gamma^+_1$ & $X^+_1$ & $L^-_2$ & $W_4$\\
Band 2 & $\Gamma^+_2$ & $\Gamma^+_2$ & $X^+_3$ & $L^-_1$ & $W_1$\\
Band 3 & $\Gamma^-_2$ & $\Gamma^-_2$ & $X^-_3$ & $L^+_1$ & $W_3$\\
Band 4 & $\Gamma^-_1$ & $\Gamma^-_1$ & $X^-_1$ & $L^+_2$ & $W_2$\\
Band 5 & $\Gamma^+_3$ & $\Gamma^+_3$ & $X^+_1$ + $X^+_3$ & $L^-_3$ & $W_1$ + $W_4$\\
Band 6 & $\Gamma^-_3$ & $\Gamma^-_3$ & $X^-_1$ + $X^-_3$ & $L^+_3$ & $W_2$ + $W_3$\\
Band 7 & $\Gamma^+_4$ & $\Gamma^+_4$ & $X^+_2$ + $X^+_5$ & $L^-_1$ + $L^-_3$ & $W_3$ + $W_5$\\
Band 8 & $\Gamma^+_5$ & $\Gamma^+_5$ & $X^+_4$ + $X^+_5$ & $L^-_2$ + $L^-_3$ & $W_2$ + $W_5$\\
Band 9 & $\Gamma^-_4$ & $\Gamma^-_4$ & $X^-_2$ + $X^-_5$ & $L^+_1$ + $L^+_3$ & $W_1$ + $W_5$\\
Band 10 & $\Gamma^-_5$ & $\Gamma^-_5$ & $X^-_4$ + $X^-_5$ & $L^+_2$ + $L^+_3$ & $W_4$ + $W_5$\\
\hline\\
\end{tabular}

\end{center}
\begin{flushleft}
Notes to Table~\ref{tab:wf_225}
\end{flushleft}
\begin{enumerate}
\item The notations of the points of symmetry in the Brillouin zone
  for $\Gamma^f_c$ follow Fig. 3.14 of Ref.~\cite{bc}, and the
  symmetry notations of the Bloch functions are defined in Table A1
  of~\cite{enio}.
\item The point groups $G_{0Ni}$ and $G_{0O}$ of the
  positions~\cite{theoriewf} of the Ni and O atoms, respectively, are
  equal to the full cubic point group $O_h$.
  The Wannier functions situated at the Ni or O atoms belong to the
  representation of $G_{0Ni}$ and $G_{0O}$, respectively, included
  below the atom. They are equal to the representations of the Bloch
  functions at point $\Gamma$ given in the next column because the
  groups $G_{0Ni}$ and $G_{0O}$ are the full cubic point group $O_h$.
\end{enumerate}
\end{table}
 

The active band is neither a magnetic nor a superconducting band
because the localized states have the full symmetry of the
paramagnetic space group $Fm3m$. I call it ``insulating band''
because it comprises all the branches crossing the Fermi level meaning
that all the electrons at the Fermi level take part in the
nonadiabatic atomic-like motion.

\begin{Definition}
  The Bloch functions of an ``insulating band'' can be unitarily
  transformed into optimally localized Wannier functions symmetry
  adapted to the full paramagnetic space group of the material in such
  a way that each Bloch function at the Fermi level belongs to the
  insulating band.
\end{Definition}

Thus, the insulating band may produce an Mott insulator in the
paramagnetic phase of NiO in the same way as the magnetic super band
may produce an insulating ground state in the antiferromagnetic
phase~\cite{enio}.

\section{Discussion}
Since both antiferromagnetic and paramagnetic NiO are insulators, the
magnetic super band of antiferromagnetic NiO~\cite{enio} as well as
the insulating band of paramagnetic NiO (Section~\ref{sec:wf}) are
(nearly) half-filled and produce a nonadiabatic atomic-like motion
evidently stabilizing the Mott insulator.  The localized states
defining the respective nonadiabatic atomic-like motion are, however,
quite different:

\subsection{Antiferromagnetic NiO}
The nonadiabatic localized states are situated at both the Ni and O
atoms and are adapted to the low symmetry of the monoclinic magnetic
group $M_9$ (defined in Equation (12) of~\cite{enio}) of the magnetic
structure. The related nonadiabatic atomic-like motion stabilizes both
the antiferromagnetic state and the Mott insulator and breaks down in
the paramagnetic phase.

\subsection{Paramagnetic NiO}
The nonadiabatic localized states are situated only at the Ni atoms
and are adapted to the full symmetry of the fcc paramagnetic phase.
They are twofold degenerate and belong to the $\Gamma_3^+$
representation of the point group $G_{0Ni}$ of their position. A
localized $\Gamma_3^+$ state may be called a ``$d$-like'' state
because it is a part of an atomic $d$ state splitting into a
$\Gamma_3^+$ and a $\Gamma_5^+$ state in the fcc crystal field, see
Table 2.7 of~\cite{bc}. The related nonadiabatic atomic-like motion
evidently stabilizes the Mott insulation because it is in accordance
with the general consent that the insulating phase occurs in crystals
with partially occupied $d$ shells~\cite{trimarchi,austin} being
``somehow localized''~\cite{brandow}. Notably, the nonadiabatic
atomic-like motion within the NHM allows a polymorphous description of
atomic-like electrons~\cite{trimarchi} because the electronic orbitals
within the nonadiabatic localized states possess the symmetry of the
lattice on the average of time, {\em but not} at any moment (while in
the adiabatic system the electrons move on rigid orbitals being
symmetric with respect to the lattice at any moment), see Section II
of~\cite{enhm}.

\section{Conclusions}
The NHM provides an atomic-like motion in both antiferromagnetic and
paramagnetic NiO being evidently the basis of the Mott insulation in
this compound. This letter demonstrates again that the nonadiabatic
atomic-like motion defined within the NHM has physical reality and
may qualify a material to be a Mott insulator. Of course, the other
antiferromagnetic and insulating metal monoxides MnO, FeO, and CoO
should also be examined in terms of magnetic super bands and
insulating bands.

\vspace{6pt} 



\funding{This publication was supported by the Open Access Publishing Fund of the University of Stuttgart.}
\acknowledgments{I am very indebted to Guido Schmitz for his support
 of my work.}

\conflictsofinterest{The author declares no conflict of interest.}


\reftitle{References}






\end{document}